\documentclass[aps,12pt,final,notitlepage,oneside,onecolumn,nobibnotes,nofootinbib,
superscriptaddress,showpacs,showkeys,centertags]{revtex4}
\usepackage{amsmath}
 \usepackage{graphicx}
 \newcommand{\ket}[1]{| {#1} \rangle}
 \newcommand{\bra}[1]{\langle {#1} |}
 \newcommand{\ave}[1]{\langle {#1} \rangle}
 \newcommand{\es}{{\mathcal E}_{\sigma}}
 \newcommand{\ep}{{\mathcal E}_{\pi}}
 \newcommand{\bp}{({\mathcal B^+}+{\mathcal B})}
 \newcommand{\bm}{({\mathcal B^+}-{\mathcal B})}
 \newcommand{\epw}{\left(\frac{1}{\ep}+\frac{1}{\omega}\right)}

\begin{document}

 \title{Anharmonic $\mathbf{O(N)}$ oscillator model in next-to-leading
        order of the $\mathbf{1/N}$ expansion}
 \author{Alan Dzhioev}
  \email[E-mail address: ]{dzhioev@thsun1.jinr.ru}
\affiliation{Bogoliubov Laboratory of Theoretical Physics, Joint
Institute for Nuclear Research, Dubna, 141980 Russia}

 \author{J. Wambach }
 \email[E-mail address: ]{wambach@physik.tu-darmstadt.de}
 \affiliation{Gesellschaft f\"ur
Schwerionenforschung, Planckstr.~1, D-64291 Darmstadt, Germany}
\affiliation{Institut f\"ur Kernphysik, Technische Universit\"at
               Darmstadt, Schlossgartenstra\ss e~9, D-64289 Darmstadt,
               Germany}

 \author{A. Vdovin }
  \email[E-mail address: ]{vdovin@thsun1.jinr.ru}
 \affiliation{Bogoliubov Laboratory of Theoretical Physics, Joint
Institute for Nuclear Research, Dubna, 141980 Russia}

 \author{Z. Aouissat}
\affiliation{Institut f\"ur Kernphysik, Technische Universit\"at
               Darmstadt, Schlossgartenstra\ss e~9, D-64289 Darmstadt,
               Germany}

 \begin{abstract}
 An extended Holstein-Primakoff mapping which incorporates both
 single- and double boson mappings is used in the context of
 $O(N+1)$ anharmonic oscillator with the global symmetry broken
 down to $O(N)$ to study the next-to-leading order of the
 $1/N$-expansion. It is shown that the Goldstone theorem is obeyed
 to that order and that the Ward identity is satisfied.
 \end{abstract}



\maketitle

\section{Introduction}
The present paper continues work started in
Refs.~\cite{Aou96,Aou97} where a boson-expansion technique, which
is widely used in nuclear structure theory, was applied to models
of quantum field theory (in particular, the linear sigma model
\cite{GelLev60}). This technique appears to be a promising tool
for building symmetry-conserving and non-perturbative approaches.
It was demonstrated that the Holstein-Primakoff mapping (HPM) is
able to systematically classify the dynamics according to the
$1/N$ expansion. For example, a $\pi\pi$ scattering equation to
lowest order in $1/N$ (which is equivalent to the random phase
approximation (RPA)) was evaluated in Ref.~\cite{Aou97} and
fulfills all required properties -- unitarity, the Ward identity
and standard RPA sum rules.

However, the perturbative boson expansion in general and the HPM
in particular rely on the bozonization of pairs of particles.
Thus, in ~\cite{Aou97} single-pion images were missing after the
mapping. Their lack proved to be an obstacle for defining
unambiguously the two-point Green's function for the Goldstone
mode. Another problem is the incorporation of single-particle
states in the HPM which is called for in finite-temperature
applications of the approach. Here, the absence of single-particle
modes causes serious difficulties in the treatment of thermal
ensembles since the single-boson (fermion) density of states plays
an important role. These shortcomings were overcome in
Ref.~\cite{Aou00} where the bosonic HPM, originally introduced
in~\cite{CurDuk89}, was extended to accommodate a single-boson
mapping. When applied to the bosonic $O(N)$ anharmonic oscillator
to leading order in the $1/N$-expansion, the extended HPM (EHPM)
was shown to render $N$ uncoupled Goldstone- as well as RPA-phonon
modes. Subsequently it was found in Ref.~\cite{ASVW01} that,
without single-boson mapping, the thermal Goldstone modes were
missing from the spectrum of the $O(N)$ model to leading order in
$1/N$.

In spite of the obvious success of the formalism its full power
should be revealed by working out the next-to-leading order since
this allows to evaluate n-point functions unambiguously.  This is
the main aim of the present work. Although the $O(N+1)$ anharmonic
oscillator is a purely quantum mechanical system its treatment in
the $1/N$ expansion, which is largely analytical, provides
valuable insight for the use of non-perturbative methods in
quantum field theory (QFT).

The paper is organized as follows. Based on the Hamiltonian of the
$O(N+1)$ anharmonic oscillator, its perturbation series in powers
of $1/\sqrt{N}$ up to next-to-leading order is constructed in
Sect.~2 within the EHPM framework. From its diagonalized form the
validity of the Goldstone theorem is inferred and the correction
to the condensate shift is computed. The validity of the Ward
identity in the same order is proved in Sect.~3. A summary and
conclusions are given in Sect.~4.

%%%%%%%%%%%%%%%%%%%%%%%%%%%%%%%%%%%%%%%%%%%%%%%
\section{Perturbative expansion of the Hamiltonian to
          order $\mathbf{1/N}$}
%%%%%%%%%%%%%%%%%%%%%%%%%%%%%%%%%%%%%%%%%%%%%%%%

 The properly scaled Hamiltonian of
the anharmonic oscillator with $O(N+1)$ symmetry broken down the
$O(N)$ group reads:
 \begin{equation}\label{ham}
  H  =  \frac{ {\vec P}_{\pi}^2}{2} +\frac{ P_{\sigma}^2}{2}
   + \frac{\omega^2}{2} \left[ {\vec X}_{\pi}^2 + X_{\sigma}^{2}\right]
    + \frac{g}{N} \left[ {\vec X}_{\pi}^2 + X_{\sigma}^2
    \right]^2 - \sqrt{N}\eta X_{\sigma}~,
\end{equation}
where we have chosen a notation reminiscent of the linear $\sigma$
model~\cite{GelLev60}, to emphasize the analogy to QFT. The
variables ${\vec X}_{\pi}$ and $X_{\sigma}$ are defined as:
\begin{equation}\label{x_oper}
 {\vec X}_{\pi}=\frac{1}{\sqrt{2\omega}}({\vec a} + {\vec
 a}^+)=\frac{1}{\sqrt{2\omega}}\sum_{i=1}^{N}(a_i+a^+_i)~,\quad
 X_{\sigma}=\frac{1}{\sqrt{2\es}}\,(b + b^+)~,
\end{equation}
and ${\vec P}_{\pi}$ and $P_{\sigma}$ denote their respective
conjugate momenta. The frequency, $\es$, of the mode $X_{\sigma}$
will be defined later.

Diagonalization of the Hamiltonian~(\ref{ham}) leads to the
excitation energies of the system. A standard approximation scheme
in many-body theory consists of splitting $H$ into a mean-field
part and a residual interaction. The mean field is obtained
selfconsistently in the Hartree-Fock (HF) or
Hartree-Fock-Bogoliubov (HFB) approximation which becomes exact in
the large-$N$ limit. It was shown in
Refs.~\cite{Aou96,Aou97,ASVW01} that, for any finite $N$, the HFB
approximation leads to a non-vanishing 'pion mass' in the exact
symmetry limit ($\eta=0$). Hence, there are no massless Goldstone
bosons in the mass spectrum. However, in the large-$N$ limit, the
Goldstone mode is recovered.  This stems from the fact that the
Fock-term is subleading and therefore vanishes. The Goldstone mode
is provided solely by the Hartree approximation. Thus, the state
of the 'Hartree pion'\footnote{We will use Goldstone mode and
'pion' interchangeably} is suitable for a perturbative treatment
of the residual interaction. In~\cite{Aou00}, such a state was
introduced in the framework of the EHPM and allows to preserve the
original single-pion degrees of freedom. The EHPM method consists
of the following mapping:
 \begin{eqnarray}\label{map}
 ({\vec{a}}{\vec{a}})_I &=&\sqrt{2N + 4(n+m)} A~,\nonumber\\
 ({\vec{a}}^{+}{\vec{a}})_I&=&2n+m~,\quad
 ({\vec{a}}^{+}{\vec{a}}^{+})_I=({\vec{a}}{\vec{a}})_I^{+}~,\nonumber\\
 (a_i)_I&=&\sqrt{2N+4(n+m)}\Gamma_N(m)\alpha_i+2\alpha_iA\Gamma_N(m)~,\nonumber\\
 (a_i^+)_I&=&(a_i)^+_I~.
 \end{eqnarray}
Here $N$ is an integer (the number of Goldstone modes), $n=A^+A$,
$m=\sum_i \alpha_i^+\alpha$, and $\Gamma_N$ is given by
 \begin{equation}
 \Gamma_N(m)=\left[\frac{m+N-2}{2(2m+N)(2m+N-2)}\right]^{1/2}~.
 \end{equation}
Thus, instead of the original pion $a_i$, we have an ideal boson
$\alpha_i$ which is a Hartree pion~\cite{Aou00}.

A vacuum state of the system $\ket{\Psi}$ is a coherent
state~\cite{Aou00} which incorporates spontaneous symmetry
breaking via the condensates for the $A$ and $b$ bosons
 \begin{equation}
 \ket{\Psi}=\exp[\ave{A}A^+ + \ave{b}b^+]\ket{0}~.
 \end{equation}
The mode $\alpha$, on the other hand, is not allowed to condense.
The state $\ket{\Psi}$ is annihilated by the shifted operators
${\mathcal A}$ and $\beta$
 \begin{equation}\label{shift}
 {\mathcal A} = A - \ave{A}~, \quad
 \beta = b - \ave{b}~,
 \end{equation}
where the condensates $\ave{A}$ and $\ave{b}$ are scaled with $N$
according to
 \begin{equation}\label{shift_v}
 \ave{A}=  d \sqrt{\frac{N}{2}}~,\quad
 \ave{b}= s\sqrt{\frac{N\es}{2}}~.
 \end{equation}
After a redefinition of  $A$ and $\beta$ a systematic $1/N$
expansion of the Hamiltonian can be performed
 \begin{equation}\label{expan}
 H=NH_0+\sqrt{N}H_1+H_2+\frac{1}{\sqrt{N}}H_3+\frac1N H_4+\cdots~.
 \end{equation}
where $H_0$ denotes the mean-field Hamiltonian. The fact that $N$
is arbitrary insures the preservation of the symmetry of $H$ at
each order. The dynamics up to order $N^{-1}$ is specified by the
$H_i$ terms in~(\ref{expan}) with $i=1,\ldots,4$ which are given
explicitly in Appendix~A.

The coherent ground state $\ket{\Psi}$ is determined by a
requirement that the ground-state energy  $NE_0\equiv\bra{\Psi}
H\ket{\Psi}/\langle\Psi|\Psi\rangle$ is minimized (note that only
$H_0$ contributes this expectation value as discussed in
Appendix~A). The minimization procedure with respect to $s$ and
$d$ then gives the following two coupled gap equations:
 \begin{eqnarray}\label{gap_eq}
 \omega^2+4gs^2+\frac{2g}{\omega}(\sqrt{1+d^2}+d)^2=\frac{\eta}{s}~,\nonumber \\
 2\omega d\sqrt{1+d^2}+\Delta(\sqrt{1+d^2}+d)^2=0~,
 \end{eqnarray}
 where the gap parameter $\Delta$ is given by
 \begin{eqnarray}
 \Delta=\frac{2gs^2}{\omega}+\frac{g}{\omega^2}(\sqrt{1+d^2}+d)^2~.\nonumber
 \end{eqnarray}
In fact, this is equivalent to the requirement that $H_1=0$.

With the gap equation~(\ref{gap_eq}), and the masses of the
Hartree pion ($\ep$) as well as the sigma boson ($\es$) (see
Appendix~A for details)
 \begin{eqnarray}\label{ep-es_mass}
 \ep^2&=&\omega^2+4gs^2+\frac{2g}{\ep}=\frac{\eta}{s}~,\nonumber\\
 \es^2&=&\omega^2+8gs^2+\frac{2g}{\ep}~,
 \end{eqnarray}
the Hartree ground-state energy and the quadratic part $H_2$ of
the Hamiltonian expansion~(\ref{expan}) can be written in a more
compact form which is particularly suitable for the further
discussions
 \begin{eqnarray}
 E_{0}=\frac{\ep^2}{8}\left[\frac{5}{\ep}-3s^2\right]-
 \frac{\es^2}{8}\left[s^2+\frac{1}{\ep}\right]-\frac{g}{4\ep^2}~,
 \end{eqnarray}
 \begin{eqnarray}\label{H2}
 H_2=\ep\sum\limits_{i}\alpha^+_i\alpha_i+\es\beta^+\beta+
 2\ep{\mathcal B^+}{\mathcal B}+\frac{g}{2\ep^2}\bp^2\nonumber\\
 +\frac{2gs}{\ep\sqrt{\es}}(\beta^++\beta)\bp+\frac{\es}{2}~.
 \end{eqnarray}
The operators $\mathcal B^+$, $\mathcal B$ appearing in (\ref{H2})
are related to $\mathcal A^+$ and $\mathcal A$ a through unitary
transformation (see~(\ref{transAB})-(\ref{coefG}) in  Appendix~A).

\begin{figure}[t]
 \centering
    \includegraphics*[width=\textwidth]{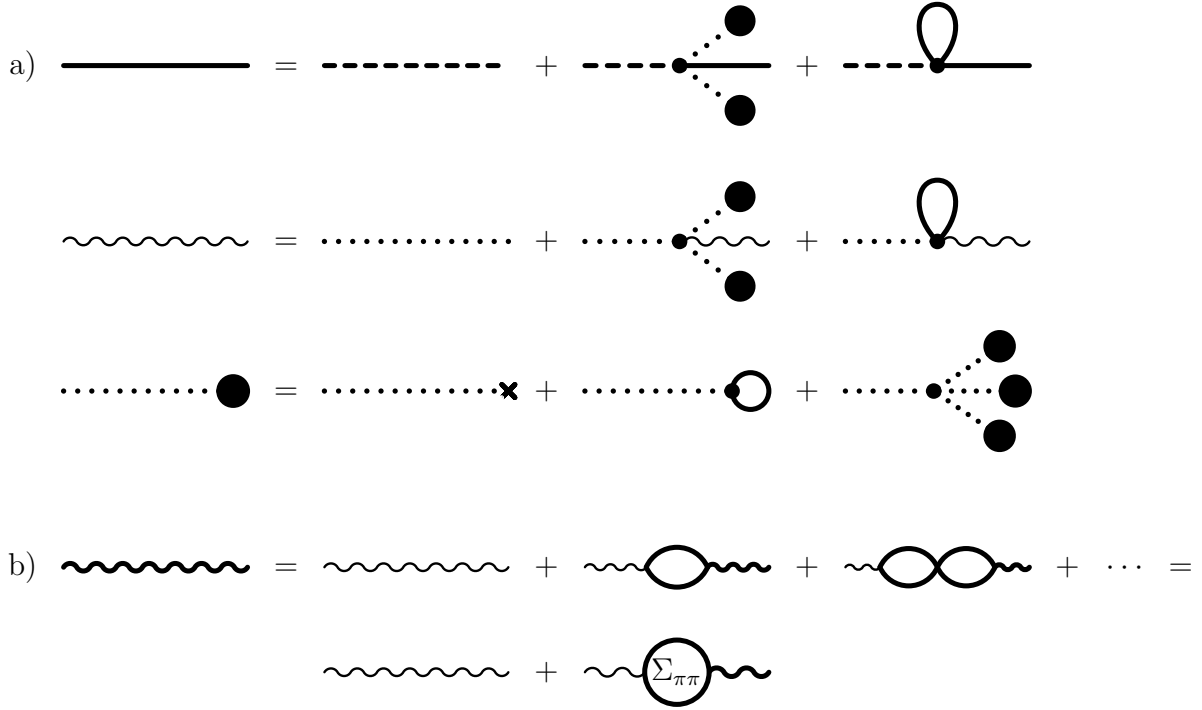}
\caption{\label{fig1} a) Diagrammatic representation of the
mean-field results for pion and sigma bosons.  The solid line
denotes the self-consistent pion propagator, the wavy line -- the
sigma propagator. The dashed (dotted) line is the two-point
Green's function of the bare $\pi$ ($\sigma$).  b) The sigma-boson
propagator in the RPA order (a thick wavy line).}
\end{figure}

The gap equation (\ref{gap_eq}) and the mean-field masses of the
Hartree pion  and  the sigma-boson~(\ref{ep-es_mass}) can be
presented diagrammatically as shown in Fig.~\ref{fig1}a.

%%%%%%%%%%%%%%%%%%%%%%%%% Figure 1 %%%%%%%%%%%%%%%%%%%%%%%%%%%%%%%%
%%%%%%%%%%%%%%%%%%%%%%%%%%%%%%%%%%%%%%%%%%%%%%%%%%%%%%%%%%%%%%%%%%%%

From Eq.~(\ref{H2}) it follows that $H_2$ describes several modes
of excitation above the Hartree ground-state energy. On the one
hand, one has independent Hartree pions which have the Goldstone
character, i.e. their frequency $\ep$ vanishes in the exact
symmetry limit ($\eta=0$), and for a finite condensate ($s\neq0$).
Besides the Goldstone modes there also exist other excitations
which can be made explicit by diagonalizing the remaining part of
$H_2$ in RPA~\cite{Aou97,Aou00}.

The next two terms, $H_3$ and $H_4$, in the Hamiltonian
expansion~(\ref{expan}) renormalize the energy of the modes
obtained after $H_2$ diagonalization and represent anharmonic
corrections. The c-numbers in $H_2$ and $H_4$ contribute to the
ground-state energy.

To diagonalize $H_2$, $H_3$ and $H_4$ we use the method described
in detail in Ref.~\cite{KM91}. We have used it
previously~\cite{Dzh04} to diagonalize  next-to-leading order
terms in the $1/N$ expansion of the Hamiltonian of the Lipkin
model \cite{Lip65}. In accordance with the prescriptions of
Ref.~\cite{KM91} the diagonalization of (\ref{expan}) within
perturbation theory can be carried out in two steps. At the first
stage RPA bosons are introduced by means of a linear canonical
transformation. This diagonalizes the sigma part of $H_2$. These
RPA bosons are then used as a basis for a systematic
diagonalization of the higher-order terms $H_3$ and $H_4$ and a
remaining part of the Hamiltonian should be written in terms of
RPA bosons. The further diagonalization can be accomplished by
means of a unitary transformation which removes all off-diagonal
terms in any order leaving the  Hamiltonian finally as a function
of a new "diagonal" bosons.

Following this procedure we define first a new RPA boson operator
$Q_{\nu}^+$ and $Q_{\nu}$ via a general Bogoliubov rotation which
mixes the sigma boson, $\beta$, and pion pairs ${\mathcal B}$:
 \begin{equation}\label{phon}
 Q_{\nu}^+=x_{\nu}\beta^++y_{\nu}\beta+
           u_{\nu}{\mathcal B^+}+v_{\nu}{\mathcal B}~.
 \end{equation}
By the equation-of-motion method~\cite{Row68} we derive RPA
equations for the amplitudes $x_{\nu},~y_{\nu},~u_{\nu},~v_{\nu}$.
The corresponding system of equations can be reduced to the
following two equations for the RPA frequencies $\Omega_{\nu}$
($\nu=1,2$) (see also Appendix~B):
\begin{eqnarray}\label{system}\left\{\begin{array}{ccl}
  \Omega_1^2+\Omega_2^2&=&4\ep^2+\es^2+{\displaystyle\frac{4g}{\ep}}~,\quad\\
  \Omega_1^2 \Omega_2^2&=&4\ep^2\left(\es^2+{\displaystyle\frac{g}{\ep}}\right)~.
                       \end{array}\right.
\end{eqnarray}
Formally, the solution of (\ref{system}) can be written as
 \begin{equation}\label{omega}
  \Omega _\nu ^2=\frac \eta s+ \frac{8gs^2}{1-
  \frac{\displaystyle 4g}
  { \displaystyle \ep \left(\Omega _\nu ^2-4\ep ^2\right)} }~.
 \end{equation}
The inverse transformation is given in Appendix~B.

Next we need in a sigma-boson propagator in RPA, to order, i.e.,
$1/N$.  It can be evaluated by solving the Dyson equation with the
mass operator taking into account pion fluctuations. The
diagrammatic representation is shown in Fig.~\ref{fig1}b. The
corresponding Dyson equation has the form
 \begin{equation}\label{dyson}
  D_{\sigma}(k^2)=\frac{i}{k^2-\es^2}+
  \frac{i}{k^2-\es^2}(-32g^2s^2){\it\Sigma_{\pi\pi}}(k^2)D_{\sigma}(k^2)~,
  \end{equation}
where ${\it\Sigma}_{\pi\pi}(k^2)$ is an infinite sum of two-pion
loop diagrams
 \begin{eqnarray}
 {\it \Sigma_{\pi\pi}}(k^2)&=&
 \int\frac{dp}{2\pi}\frac{i^2}{(p^2-\ep^2)([p-k]^2-\ep^2)}\nonumber\\&&
 +(-4ig)\left[\int\frac{dp}{2\pi}\frac{i^2}{(p^2-\ep^2)([p-k]^2-\ep^2)}\right]^2+
 \cdots\nonumber\\&=&
 \frac{i}{\ep(k^2-4\ep^2)}+(-4ig)\left[\frac{i}{\ep(k^2-4\ep^2)}\right]^2+\cdots\nonumber\\&=&
 \frac{i}{\ep(k^2-4\ep^2-4g/\ep)}~.
 \end{eqnarray}
The additional factors $1/N$ at each vertex are precisely canceled
by the factor $N$ appearing due to summing over $N$ internal pions
in each bubble. Inserting the explicit form of ${\it
\Sigma_{\pi\pi}}(k^2)$ into Eq.~(\ref{dyson}) one obtains
 \begin{equation}
 D_{\sigma}(k^2)=\frac{i}
 {k^2-\es^2+32i g^2s^2 {\it\Sigma}_{\pi\pi}(k^2)}
 \end{equation}
Now the RPA frequencies  $\Omega_{\nu}$ can be obtained by solving
$[D_\sigma(\Omega_{\nu}^2)]^{-1}=0$. They are identical to those
in Eq.~(\ref{omega}) which were obtained by using the
equation-of-motion method. In terms of $\Omega_{1,2}$ the
sigma-boson propagator reads
 \begin{eqnarray}
 D_{\sigma}(k^2)=i\frac{k^2-4\ep^2-4g/\ep}{(k^2-\Omega_1^2)(k^2-\Omega_2^2)}~.
 \end{eqnarray}
It should be mentioned that in this approximation the condensate
$s$ does not receive any contributions from RPA fluctuations.
Moreover, if $k^2\to\Omega^2_{\nu}$ the sigma-boson propagator
becomes
 \begin{eqnarray}
  D_{\sigma}(k^2)\sim\frac{iZ_{\nu}}{k^2-\Omega_{\nu}^2}~,
  \end{eqnarray}
where the field renormalization constants $Z_{\nu}$ are given by
 \begin{equation}
 Z_{\nu}=(\es^2-\Omega_{\nu'}^2)/(\Omega_{\nu}^2-\Omega_{\nu'}^2)=\phi_{\nu'}^2~.
 \end{equation}
 Here and hereafter the notation $\nu'$ means that $\nu'\ne\nu$.

Now, the  $H_2$ part of the Hamiltonian can be written in diagonal
form in terms of Hartree pions and RPA bosons
 \begin{equation}\label{H2RPA}
 H_2=\ep\sum\limits_{i=1}^{N}\alpha^+_i\alpha_i+
 \sum\limits_{i=1,\,2}\Omega_{\nu}Q^+_{\nu}Q_{\nu}+
 \frac{\Omega_1+\Omega_2}{2}-\ep~.
 \end{equation}
 The last two (c-number) terms  in (\ref{H2RPA}) represent the RPA corrections
 to the Hartree ground-state energy $NE_0$.

Having diagonalized $H_2$ we now turn to $H_3$ and $H_4$ and
express them in terms of the RPA phonons. The explicit expressions
are given in Appendix B. To diagonalize  we use the quantum analog
of the classical Birkhoff-Gustavson approach~\cite{Mar82,BG} as
was done in~\cite{Dzh04}. The diagonal terms in $H_4$ (Appendix B)
provide anharmonic corrections of order $N^{-1}$ to RPA. On the
other hand, all terms in $H_3$ include only odd powers of
$Q_{\nu},~Q^+_{\nu}$ and have, therefore, vanishing diagonal
matrix elements in the RPA basis. They first contribute in
second-order perturbation theory and are thus of the same order as
the diagonal contributions of $H_4$. Therefore, these parts of the
Hamiltonian must be considered together.

As has been mentioned, it is possible to define a perturbative,
unitary transformation which removes all off-diagonal terms in any
given order of the $1/N$- expansion. At the order of interest,
this transformation has to be chosen such as to eliminate the
$H_3$ part of the Hamiltonian expansion~(\ref{expan}), thereby
shifting their effects to renormalized diagonal terms of $H_4$.

To this end, we define new "diagonal" pion and sigma operators
$(\alpha_i^+)_d$ and $(Q_{\nu}^+)_d$ which are obtained via a
unitary transformation from the $\alpha_i$ and $Q_{\nu}$ operators
such that
\begin{eqnarray}\label{diag_phon}
 (\alpha_i^+)_d=e^{-iS/\sqrt{N}}\alpha_i^+e^{iS/\sqrt{N}}~,~\quad
 (Q_{\nu}^+)_d=e^{-iS/\sqrt{N}}Q_{\nu}^+e^{iS/\sqrt{N}}~.
 \end{eqnarray}
The Hamiltonian is then rewritten in terms of these "diagonal"
operators\footnote{Here we have adopted the following notation.
For any operator $F$, depending, e.g., on $O$ and $O^+$, $F_d$ is
obtained by the replacement $O\to O_d$ and $O^+\to O^+_d$, i.e.
$F=e^{iS_d}F_de^{-iS_d}$. Of course, one has $e^{iS}=e^{iS_d}$. We
also use the following appreciation for a product of operators:
$a_db_d\ldots z_d\equiv(ab\ldots z)_d$.} as:
  \begin{eqnarray}
  H=e^{-iS_d/\sqrt{N}} H_d e^{iS_d/\sqrt{N}}=
  H_d+\frac{1}{\sqrt N}[H, iS]_d+\frac{1}{2N}[[H, iS],
  iS]_d+\cdots~.\nonumber
  \end{eqnarray}
Using the Hamiltonian expansion~(\ref{expan}), we obtain after
some algebra, in the order of interest
  \begin{eqnarray}\label{hdiag}
  H&=&NH_0+(H_2)_d+\frac{1}{\sqrt N}\left\{[H_2,iS]+ H_3\right\}_d
  \nonumber\\&&
  +\frac{1}{2N}\left\{\left[[H_2,iS],
  iS\right]+2[H_3, iS]+2H_4\right\}_d+\cdots~.
  \end{eqnarray}
 Therefore, the condition for eliminating the off-diagonal terms of
 order of $1/\sqrt{N}$ can be easily read off as
  \begin{eqnarray}\label{cond}
  [H_2,iS]_d=-(H_3)_d~.
  \end{eqnarray}
and completely determines the operator $iS_d$. Using this result
in Eq.~(\ref{hdiag}) we obtain after some algebra (Appendix~C) the
full diagonal part of the sigma-model Hamiltonian in
next-to-leading order of the $1/N$ expansion:
 \begin{eqnarray}\label{Q-ham}
 H&=&N(H_0+\frac{\Omega_1+\Omega_2-2\ep}{2N})+
 (\ep+\frac1N\ep')\sum_i(\alpha^+_i\alpha_i)_d
 \nonumber\\&&
 +\sum_{\nu}(\Omega_{\nu}+\frac1N\Omega'_{\nu})(Q^+_{\nu}Q_{\nu})_d
 +\frac{1}{N}\Bigl\{\frac{1}{2}\sum_{i,j}V_{\pi\pi\to\pi\pi}\alpha^+_i\alpha^+_j\alpha_j
 \alpha_i\nonumber\\&&
 +\sum_{i,\nu}V_{\pi\sigma_\nu\to\pi\sigma_\nu}\alpha^+_iQ^+_{\nu}Q_{\nu}\alpha_i
 +\frac{1}{2}\sum_{\nu,\nu'}V_{\sigma_{\nu}\sigma_{\nu'}\to\sigma_{\nu}\sigma_{\nu'}}
 Q^+_{\nu}Q^+_{\nu'}Q_{\nu'}Q_{\nu}\Bigl\}_d~.
 \end{eqnarray}
Besides RPA corrections to the Hartree ground-state energy it
contains all anharmonic corrections caused by residual
interactions between different modes as well as higher-order
contributions $\ep'$,~$\Omega_{\nu}'$ to the masses of pion- and
sigma-bosons, respectively. As can be seen, there are three
sources of residual interaction: the pion-pion term $\sim
V_{\pi\pi\to\pi\pi}$, the pion-sigma term $\sim
V_{\pi\sigma_\nu\to\pi\sigma_\nu}$, and the sigma-sigma term $\sim
V_{\sigma_{\nu}\sigma_{\nu'}\to\sigma_{\nu}\sigma_{\nu'}}$. The
explicit form of these couplings  as well as of $\ep'$ and
$\Omega_{\nu}'$ are given in Appendix~C. It is evident from
(\ref{pion_pr}) and (\ref{matr_el}) that the $1/N$ order
correction $\ep'$ to the pion mass and pion-pion interaction term
$V_{\pi\pi\to\pi\pi}$ vanish in the exact symmetry limit $\eta=0$.
This implies that the pion keeps its Goldstone character in
next-to-leading order of the $1/N$ expansion.

Next-to-leading order corrections also contribute to the
sigma-boson shift (\ref{shift_v}). The corresponding correction
reads
 \[
 s'=\frac{1}{\sqrt{2N\es}}\ave{\beta^++\beta}~,
 \]
where $\ave{\ldots}$ denotes the expectation value with respect to
the vacuum of "diagonal" bosons. Expressing $(\beta^++\beta)$
through these bosons we obtain
 \begin{equation}\label{shift_cor}
 s'=-\frac{2s\ep'}{N\ep}+\frac{8gs}{N}\left(\frac{\ep}{\Omega_1^2\Omega_2^2}+
 \frac{3}{\Omega_1(\Omega_1^2-\Omega_2^2)}-
 \frac{3}{\Omega_2(\Omega_1^2-\Omega_2^2)}\right)~.
 \end{equation}

%%%%%%%%%%%%%%%%%%%%%%%%%%%%%%%%%%%%%%%%%%%%%%%
\section{Ward identity in next-to-leading order}
%%%%%%%%%%%%%%%%%%%%%%%%%%%%%%%%%%%%%%%%%%%%%%%%

Aside from the Goldstone character of the pion the well-known Ward
identity relating the pion propagator to the sigma-boson
shift~\cite{Lee} should also remain valid. In this Section we
prove that the Ward identity is indeed fulfilled in
next-to-leading order of the $1/N$ expansion.

\begin{figure}[t]
 \centering
 \includegraphics*[width=\textwidth]{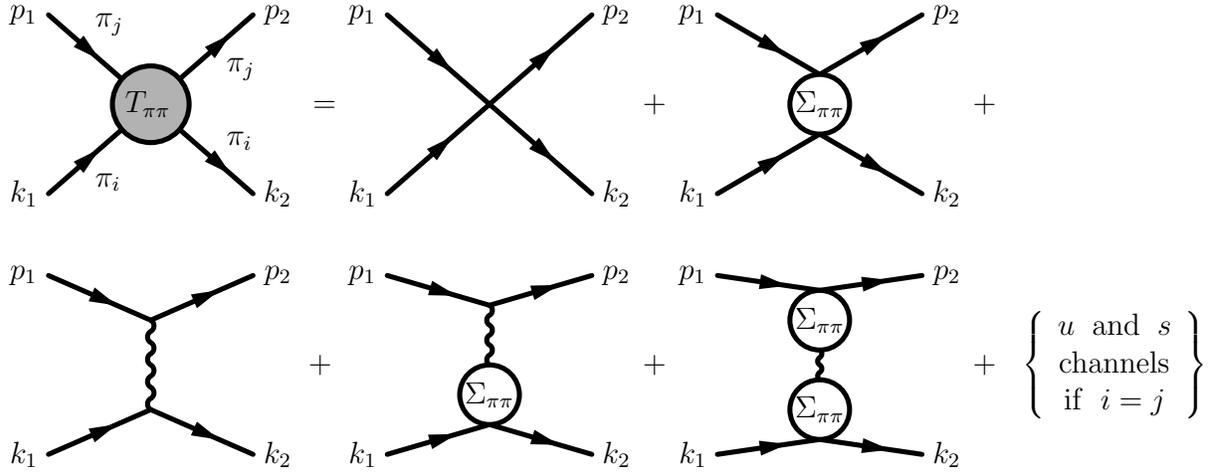}
 \caption{\label{fig2} Diagrammatic representation of $\pi$--$\pi$ scattering.
 The wavy line denotes the sigma boson RPA propagator while
 ${\it\Sigma}_{\pi\pi}$ contains an infinite sum of interacting
 pion loops.}
\end{figure}

The Ward identity means that the pion propagator with $1/N$
corrections to the pion self-energy and  the sigma-boson shift
value calculated in the same order should satisfy the following
relation:
  \begin{equation}\label{ward}
  \left[iD_{\pi}(0)\right]^{-1}=\frac{\eta}{s+N^{-1}s'}~.
  \end{equation}
To prove Eq.~(\ref{ward}) we will use a diagrammatic technique to
obtain the pion propagator in the order of interest. A necessary
step is to calculate the $\pi\pi$-scattering $T$-matrix which
contributes to the pion self-energy at order $1/N$. There exist
two types of $\pi\pi$-interactions: those involving contact
interactions (first two diagrams in Fig.~\ref{fig2}) and those
arizing from sigma-boson exchange (remaining diagrams). It is
important to stress that, in the $1/N$ expansion, diagrams of the
same topological structure can contribute in different orders. If
$i\neq j$ in Fig.~\ref{fig2}, then there is no summation over an
internal pion in the $s$- and $u$-channels. Adding a new pion
bubble increases the order of the diagram by $1/N$. Therefore, in
this case an interaction exists only in $t$-channel to leading
order. On the other hand, in the case  $i=j$ all the three
channels $s,~t,~u$ contribute to the scattering process.

%%%%%%%%%%%%%%%%%%%%%%%%%Figure 2%%%%%%%%%%%%%%%%%%%%%%%%%%%%%%%%%%%%%
%%%%%%%%%%%%%%%%%%%%%%%%%%%%%%%%%%%%%%%%%%%%%%%%%%%%%%%%%%%%%%%%%%%%%%%%

The first two diagrams give the following contribution to the
$T$-matrix:
 \begin{eqnarray}
 T_{\rm cont}(l^2)&=&
 8g\left\{1+(-4ig){\it\Sigma}_{\pi\pi}(l^2)\right\}=8g\frac{l^2-4\ep^2}{l^2-4\ep^2-4g/\ep}~,
 \end{eqnarray}
 where $\vec l=\vec k_1-\vec k_2$. A given channel is specified by a fixed value
 of $l^2$: $t$-channel $l^2=0$, $s$-, $u$-channel $l^2=4\ep^2$. From the
 $\sigma-$exchange
 diagrams we have
 \begin{eqnarray}
  T_{\rm exch}(l^2)&=&-56g^2s^2iD_{\sigma}(l^2)
 \left\{1+(-4ig){\it\Sigma}_{\pi\pi}(l^2)\right\}^2 \nonumber\\&=&
 \frac{56g^2s^2(l^2-4\ep^2)^2}{(l^2-4\ep^2-4g/\ep)(l^2-\Omega_1^2)(l^2-
 \Omega_2^2)}~.
 \end{eqnarray}
 Thus the full $T$ matrix in a given channel reads
 \begin{eqnarray}\label{t_matrix}
 T_{\pi\pi}(l^2)&=&T_{\rm cont}(l^2)+T_{\rm exch}(l^2)=
 8g\frac{(l^2-4\ep^2)(l^2-\ep^2)}{(l^2-\Omega_1^2)(l^2-\Omega_2^2)}~.
 \end{eqnarray}

As was already mentioned,  if  $i\neq j$,  only the interaction in
the $t$-channel should be taken into account, hence $l^2=0$. Then
 \[
 T_{\pi_i\pi_j}=T_{\pi\pi}(0)=\frac{32g\ep^4}{\Omega_1^2\Omega_2^2}~.
 \]
On the other hand, if $i=j$,
 \[
 T_{\pi_i\pi_i}=2T_{\pi\pi}(0)+T_{\pi\pi}(4\ep^2)=\frac{64g\ep^4}
 {\Omega_1^2\Omega_2^2}~.
 \]
Dividing the $T$ matrix by a factor $4\ep^2(1+\delta_{ij})$ which
arises from the relativistic normalization condition, we obtain
the matrix element of the pion-pion interaction
$V_{\pi\pi\to\pi\pi}$~(\ref{matr_el}) calculated in the previous
section in the framework of the EHPM.

\begin{figure}[t]
 \centering
  \includegraphics*[width=\textwidth]{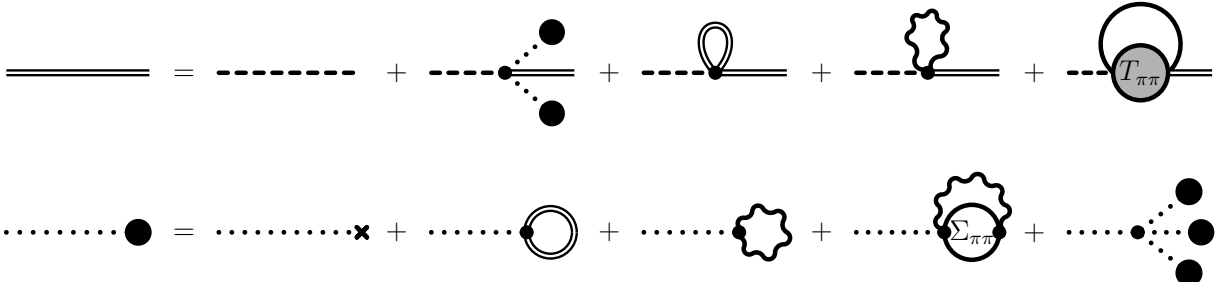}
  \caption{\label{fig3}
Upper part: Dyson equation for the pion propagator (doubled line)
to order $1/N$. Lower part: Dyson  equation for the sigma-boson
condensate value in the order $1/N$.}
 \end{figure}

We are now able to calculate the pion propagator to order $1/N$.
The  diagrammatic representation of Dyson equation and the
condensate is shown in Fig.~\ref{fig3}.

%%%%%%%%%%%%%%%%%%%%%%%%%Figure 3%%%%%%%%%%%%%%%%%%%%%%%%%%%%%%%%%%%%%%%
%%%%%%%%%%%%%%%%%%%%%%%%%%%%%%%%%%%%%%%%%%%%%%%%%%%%%%%%%%%%%%%%%%%%%%%%

As one can see there appear two additional diagrams as compared to
Fig.~\ref{fig1}, involving a sigma bubble and $T_{\pi\pi}$. Thus,
to order $1/N$, we obtain two coupled equations for the pion
propagator $D_{\pi}(k^2)$ and the sigma-boson shift  $s+N^{-1}s'$
 \begin{eqnarray}\label{pion_prop}
 [iD_{\pi}(k^2)]^{-1}&=&-k^2+\omega^2+4g(s+N^{-1}s')^2+4g\int\frac{dp}{2\pi}
 D_{\pi}(p^2)\nonumber\\&&
 +\frac{4g}{N}\int\frac{dp}{2\pi}D_{\sigma}(p^2)+
 \frac{8g}{N}\int\frac{dp}{2\pi}\frac{i}{p^2-\ep^2}T_{\pi\pi}([k-p]^2)~,
 \end{eqnarray}
 \begin{eqnarray}\label{shift_eq}
 (s+N^{-1}s')&=&\frac{\eta}{\omega^2}-\frac{4g}{\omega^2}(s+N^{-1}s')\int
 \frac{dp}{2\pi}D_{\pi}(p^2)\nonumber\\&&-
 \frac{4g}{N\omega^2}(s+N^{-1}s')\int\frac{dp}{2\pi}D_{\sigma}(p^2)
 \left\{3+2(-4ig){\it\Sigma}_{\pi\pi}(p^2)\right\}\nonumber\\&&-
 \frac{4g}{\omega^2}(s+N^{-1}s')^3~,
 \end{eqnarray}
where $D_{\sigma}(p^2)$ is the sigma-boson propagator in the RPA
order. Except for terms involving the pion propagator
$D_{\pi}(p^2)$ all other pieces in~(\ref{pion_prop})
and~(\ref{shift_eq}) can be calculated analytically. After some
algebra we obtain
 \begin{eqnarray}
 \frac{\eta}{s+N^{-1}s'}&=&\omega^2+4g(s+N^{-1}s')^2+4g\int\frac{dp}{2\pi}D_{\pi}
 (p^2)\nonumber\\&&-
 \frac{2g}{N}\left[\frac{\Omega_2^2-\es^2}{\Omega_1(\Omega_1^2-\Omega_2^2)}-
 \frac{\Omega_1^2-\es^2}{\Omega_2(\Omega_1^2-\Omega_2^2)}\right]\nonumber\\&&-
 \frac{4g}{N}\left[\frac{4\ep^2-\Omega_1^2}{\Omega_1(\Omega_1^2-\Omega_2^2)}-
 \frac{4\ep^2-\Omega_2^2}{\Omega_2(\Omega_1^2-\Omega_2^2)}\right]=
                   [iD_\pi(0)]^{-1}~.
 \end{eqnarray}
This proves the Ward identity~(\ref{ward}) in next-to-leading
order in $1/N$ expansion.

We show that  the condensate $s'$ in Eq.~(\ref{shift_eq})  are
equal to that in Eq.~(\ref{shift_cor}) which is calculated with
the EHPM technique. To do so we should express the pion propagator
(\ref{pion_prop}) in terms of the pion mass in next-to leading
order, $(\ep+N^{-1}\ep')$, which is the solution of
 \[
   [D_{\pi}(k^2)]^{-1}=0~.
 \]
In terms of this mass, the pion propagator reads
  \begin{eqnarray}
 [iD_{\pi}(k^2)]^{-1}&=&-k^2+(\ep+N^{-1}\ep')^2\nonumber\\&&-
 \frac{8g}{N}\int\frac{dp}{2\pi}\frac{i}{p^2-\ep^2}\left[T_{\pi\pi}([\ep-p]^2)-T_
 {\pi\pi}([k-p]^2)\right]~.
  \end{eqnarray}
Putting $k^2=0$ we obtain
  \begin{eqnarray}
  [iD_\pi(0)]^{-1}&=&(\ep+N^{-1}\ep')^2\nonumber\\&&-
  \frac{8g}{N}\int\frac{dp}{2\pi}\frac{i}{p^2-\ep^2}\left[T_{\pi\pi}([\ep-p]^2)-T_
  {\pi\pi}(p^2)\right]\nonumber\\&=&
  (\ep+N^{-1}\ep')^2\nonumber\\&&-
  \frac{8g\ep^2}{N}\left[\frac{\ep}{\Omega_1^2\Omega_2^2}+
  \frac{3}{\Omega_1(\Omega_1^2-\Omega_2^2)}-\frac{3}{\Omega_2(\Omega_1^2-
  \Omega_2^2)}\right]~.
  \end{eqnarray}
Using this result in the lhs. of Eq.~(\ref{ward}), one recovers
the result in Eq.~(\ref{shift_cor}).

%%%%%%%%%%%%%%%%%%%%%%%%%%%%%%%%%%%%%%%%%
\section{Summary and conclusions}
%%%%%%%%%%%%%%%%%%%%%%%%%%%%%%%%%%%%%%%%
In this paper we have applied the extended Holstein-Primakoff
boson mapping~\cite{Aou00} to the bosonic $O(N)$ anharmonic
oscillator which mimics some essential features of the linear
sigma model of QFT. The approach that we use fundamentally relies
on the bozonization of the generators of the $Sp_2\otimes N_1$
semidirect product group and is very appealing from many aspects
and particularly from its symmetry-conserving character. The
extended boson HP mapping exploits both single-boson and two-boson
images. The former seems to be of crucial importance to correctly
embed the correct quantum statistics in the mapping~\cite{ASVW01}.
Moreover, it permits unambiguously to define the two-point
function for the Goldstone mode.

In Refs.~\cite{Aou00,ASVW01}, the $O(N)$ anharmonic oscillator has
been treated within the EHPM formalism to leading-order in the
$1/N$ expansion which is equivalent to the  Hartree + RPA scheme.
In that approximation the system  appears as a system of $N$
Goldstone modes (pions) and two RPA modes which are superposition
of the initial sigma-boson and auxiliary "two-pion" modes.

In next-to-leading order the above excitations are coupled with
each other. We have calculated the corresponding coupling matrix
elements \footnote{For the present, purely quantum-mechanical
model, these matrix elements play a role of four-point Green's
functions for QFT in four space-time dimensions.}. Furthermore,
next-to-leading order corrections to pion- and sigma-  masses as
well as to the condensate shift were obtained through perturbative
diagonalization of the Hamiltonian using single-pion and RPA
bosons as a basis. We have proved that in next-to-leading order
the Goldstone theorem remains valid and the Ward identity is
satisfied. These results constitute an encouraging step towards
the application of the EHPM techniques to field theoretical modes,
such as the linear sigma-model.

%%%%%%%%%%%%%%%%%%%%%%%%%%%%%%%%%
%\renewcommand{\thesection}{}
\section*{Acknowledgments}
%%%%%%%%%%%%%%%%%%%%%%%%%%%%%%%%%
The authors are grateful to Dr.~A.~Storozhenko for his essential
contribution at the early stages of these studies. The work was
supported in part by the Heisenberg-Landau Program and by the
Deutsche Forschungsgemeinschaft through the SFB 634.

%%%%%%%%%%%%%%%%%%%%%%%%%%%%%%%%%%%%%%%%%
\appendix
\section{}
%%%%%%%%%%%%%%%%%%%%%%%%%%%%%%%%%%%%%%%%%

In this Appendix the boson expansion series up to next-to-leading
order in $1/N$ for the Hamiltonian (1) is presented.

In terms of the shifted boson operators $\beta$ and $\mathcal A$
the $X_{\sigma}^2$ and ${\vec X}_{\pi}^2$ take the following form:
\begin{eqnarray}\label{root}
 X_{\sigma}^2&=&
 Ns^2+\sqrt{N}s\sqrt{\frac{2}{\es}}(\beta^++\beta)+
 \frac{1}{2\es}(\beta^++\beta)^2~,\nonumber\\
  {\vec X}_{\pi}^2&=&\frac{1}{2\omega}\Biggl\{N(1+2d^2)/2+
  \sqrt{2N}d({\mathcal A}^++{\mathcal A})+
  2{\mathcal A}^+{\mathcal A}+m\nonumber\\&&+
  \sqrt{2NK}({\mathcal A}^++d\sqrt{N/2})\times\nonumber\\&&
  \left[1+\frac{2m+2{\mathcal A}^+{\mathcal A}+\sqrt{2N}d({\mathcal A}^++{\mathcal A})}
  {NK}\right]^{1/2}+{\rm h.c.}\Biggr\}~,
  \end{eqnarray}
where $K=1+d^2$. After a Taylor expansion of a square root
in~(\ref{root}) up to order $N^{-1}$, the operator ${\vec
X}_{\pi}^2$ is given by
 \begin{eqnarray}
  {\vec X}_{\pi}^2&=&
  N\frac{1}{4\omega}(\sqrt{K}+d)^2+
  N^{1/2}\frac{(\sqrt{K}+d)^2}{2\omega\sqrt{2K}}({\mathcal A}^++{\mathcal A})
  \nonumber\\&&+
 \frac{\sqrt{K}+d}{8\sqrt{K}\omega}
 \left[({\mathcal A}^++{\mathcal A})^2\frac{1}{K}-({\mathcal A}^+-{\mathcal
 A})^2+4(m-1)\right]
 \nonumber\\&&+
 \frac{(\sqrt{K}+d)^2}{8\omega}\left[({\mathcal A}^++{\mathcal A})^2\frac{1}{K}-
 ({\mathcal A}^+-{\mathcal A})^2(\sqrt{K}-d)^2\right]
 \nonumber\\&&+
 N^{-1/2}\frac{({\mathcal A}^++{\mathcal A})}{8\sqrt{2}\omega K^{3/2}}\left[
 ({\mathcal A}^++{\mathcal A})^2\frac{1}{K}-({\mathcal A}^+-{\mathcal
 A})^2+4(m-1)\right]
 \nonumber\\&&-
 N^{-1}\frac{d}{64K^{3/2}\omega}\Biggl\{
 \left[5({\mathcal A}^++{\mathcal A})^2\frac{1}{K}-({\mathcal A}^+-{\mathcal A})^2+4(m-1)\right]
 \times\nonumber\\&&
 \left[({\mathcal A}^++{\mathcal A})^2\frac{1}{K}-({\mathcal A}^+-{\mathcal
 A})^2+4(m-1)\right]
 -\frac{4(d^2-1)}{K}\Biggr\}+{\rm h.c.}
  \end{eqnarray}
Substituting this expression into Eq.~(\ref{expan})  leads to the
required $1/N$ expansion of the model Hamiltonian. The first term
$NH_0$ yields the ground-state energy in the Hartree approximation
 \begin{eqnarray}\label{Hartree}
 E_0\equiv\frac{\bra{\Psi}
 H\ket{\Psi}}{N\langle\Psi|\Psi\rangle}
 &=&\frac{\omega}{2}(1+2d^2)+
 \left(\frac{\omega^2s^2}{2}-\eta s\right)\nonumber\\&&+
 g\left(\frac{(\sqrt{K}+d)^4}{4\omega^2}+s^4+
 \frac{s^2(\sqrt{K}+d)^2}{\omega}\right)~.
 \end{eqnarray}
Minimizing $E_0$ determines the values of $d$ and $s$ as the roots
of the coupled gap equations~(\ref{gap_eq}). The next term $H_1$
is linear with respect to $\mathcal A$ and $\beta$:
 \begin{eqnarray}
 H_1&=&\frac{(\mathcal A^++\mathcal A)}{\sqrt{2K}}
 \left[2\omega d\sqrt{K}+\Delta(\sqrt{K}+d)^2\right]\nonumber\\&&+
 \frac{(\beta^++\beta)s}{\sqrt{2\es}}\left[\omega^2+2\Delta\omega-\frac{\eta}{s}
 \right]
 \end{eqnarray}
and vanishes at the minimum of $E_0$.

For the following it is particularly suitable to define a new
operators ${\mathcal B^+}$ and ${\mathcal B}$
\begin{equation}\label{transAB}
 {\mathcal A^+}=G_+{\mathcal B^+}+G_-{\mathcal B}~.
\end{equation}
In order for the $\mathcal B^+$, $\mathcal B$ operators to obey
the same algebra as $\mathcal A^+$, $\mathcal A$ the coefficients
$G_+$ and $G_-$ have to be chosen such that $ G_+^2-G_-^2=1$.
 Explicitly, they are given by
 \begin{equation}\label{coefG}
 G_\pm=\frac12\left[\sqrt{K}\pm\frac{1}{\sqrt{K}}\right]
 \end{equation}
and we have
 \begin{eqnarray}
 ({\mathcal A^+}+{\mathcal A})=\bp\sqrt{K}~,&&
  ({\mathcal A^+}-{\mathcal A})=\bm/\sqrt{K}~.
 \end{eqnarray}

In terms of $\mathcal B^+$ and ${\mathcal B}$ the quadratic part
of $H$ now reads as
 \begin{eqnarray}
 H_2&=&\left[\omega+\Delta\frac{\sqrt{K}+d}{\sqrt{K}}\right]
 \sum\limits_{i}\alpha_i^+\alpha_i
   \nonumber\\&&+
 \left[\frac{\omega^2-\es^2}{4\es}+\frac{\omega\Delta}{2\es}+
       \frac{2gs^2}{\es}\right](\beta^+\beta^++\beta\beta)
   \nonumber\\&&+
 \left[\frac{\omega^2+\es^2}{4\es}+\frac{\omega\Delta}{2\es}+
       \frac{2gs^2}{\es}\right](2\beta^+\beta+1)
   \nonumber\\&&+
 \frac{g(\sqrt{K}+d)^4}{2\omega^2}\bp^2+
 \frac{2\omega}{(\sqrt{K}+d)^2}{\mathcal B^+}{\mathcal B}\nonumber\\&&+
 \frac{2gs(\sqrt{K}+d)^2}{\omega\sqrt{\es}}(\beta^++\beta)\bp~.\nonumber
 \end{eqnarray}
 From this expression, and more precisely from the coefficient
 in front of the term
 $\sum_i \alpha^+_i\alpha_i$, one can deduce the existence of $N$
 uncoupled modes that are nothing but the Hartree pions with mass
 $\ep$ given by
 \begin{eqnarray}\label{ep}
 \ep&=&\omega+\Delta\frac{\sqrt{K}+d}{\sqrt{K}}=
 \frac{\omega}{(\sqrt{K}+d)^2}~,\nonumber\\
 \ep^2&=&\omega(\omega+2\Delta)=\omega^2+4gs^2+\frac{2g}{\ep}~.
 \end{eqnarray}
Furthermore, demanding that the bilinear part of $H_2$ in the
$\beta$ operators is diagonal, i.e.
 \[
 \frac{\omega^2-\es^2}{4\es}+\frac{\omega\Delta}{2\es}+
       \frac{2gs^2}{\es}=0~,\nonumber
 \]
we obtain the sigma-boson mass $\es$
 \begin{equation}\label{es}
 \es^2=
 \omega^2+12gs^2+\frac{2g}{\ep}=\ep^2+8gs^2~.
 \end{equation}
With these definitions of $\ep$ and $\es$ the quadratic part $H_2$
of the Hamiltonian can be recast in the compact form given in
Eq.~(\ref{H2}).

The next-to-leading order terms $H_3$ and $H_4$ in the Hamiltonian
expansion look as follows:
\begin{eqnarray}\label{H3}
 H_3&=&\frac{1}{4\sqrt{2K}}\left\{-\ep
 d\bp+\frac{\sqrt{K}-d}{\ep}\left(
 \frac{g}{\ep}\bp+\frac{2gs}{\sqrt{\es}}(\beta^++\beta)\right)\right\}
 \nonumber\\&&\times
 \left\{4(m-1)+\bp^2-\bm^2\frac1K\right\}\nonumber\\&&+
 \frac{1}{4\sqrt{2}}\left\{
 \frac{g}{\ep}\bp+\frac{2gs}{\sqrt{\es}}(\beta^++\beta)\right\}
 \nonumber\\&&\times
 \left\{\frac{1}{\ep}\bp^2-
 \frac{1}{K\omega}\bm^2+ \frac{2}{\es}(\beta^++\beta)^2\right\}+{\rm
 h.c.}~,
 \end{eqnarray}
 \begin{eqnarray}\label{H4}
 H_4&=&\frac1K\left(\ep
 d^2+\frac{g}{\ep\omega}\right)(m-1)^2\nonumber\\&&+
 \frac{1}{4K}\left\{
 2\left(\ep d^2+\frac{g}{\ep\omega}\right)
 \left(3\bp^2-\frac{1}{K}\bm^2\right)\right.\nonumber\\&&+
 \frac{8gs}{\sqrt{\es}\omega}\bp(\beta^++\beta)+
 g\epw\nonumber\\&&\left.\times
 \left(\frac{1}{\ep}\bp^2-\frac{1}{K\omega}\bm^2
 +\frac{2}{\es}(\beta^++\beta)^2\right)\right\}
 (m-1)\nonumber\\&&+
 \frac{1}{32K}\left\{
 \left(\ep d^2+\frac{g}{\ep\omega}\right)
 \left(5\bp^2-\frac{1}{K}\bm^2\right)\right.\nonumber\\&&+
 \frac{8gs}{\sqrt{\es}\omega}\bp(\beta^++\beta)+
 g\epw\nonumber\\&&\left.\times
 \left(\frac{1}{\ep}\bp^2-
 \frac{1}{K\omega}\bm^2+\frac{2}{\es}(\beta^++\beta)^2\right)\right\}\nonumber\\&&\times
 \left\{\bp^2-\frac{1}{K}\bm^2\right\}\nonumber\\&&
 +\frac{g}{32}\left\{\frac{1}{\ep}\bp^2-
 \frac{1}{K\omega}\bm^2+\frac{2}{\es}(\beta^++\beta)^2\right\}^2\nonumber\\&&+
 \frac{g}{4K^2\ep\omega}-\frac{\ep d^2(d^2-1)}{8K^2}+{\rm
 h.c.}
 \end{eqnarray}
 %

%%%%%%%%%%%%%%%%%%%%%%%%%%%%%%%%%%%%%%%%%
\section{}
%\renewcommand{\theequation}{B.\arabic{equation}}
%\setcounter{equation}{0}
%%%%%%%%%%%%%%%%%%%%%%%%%%%%%%%%%%%%%%%%%
In this Appendix some details concerning the Bogoliubov rotation
(\ref{phon}) are presented.

We start with the following definition of the RPA excitation
operator:
 \[
 Q_{\nu}^+=x_{\nu}\beta^++y_{\nu}\beta+u_{\nu}{\mathcal B^+}+v_{\nu}{\mathcal
 B}~.
\]
 Using the equation of motion~\cite{Row68}
 \begin{equation}
  \bra{{\rm RPA}}[\delta Q_{\nu},[H_2,Q^+_{\nu}]]\ket{{\rm RPA}}=
  \Omega_{\nu} \bra{{\rm RPA}}[\delta Q_{\nu}, Q^+_{\nu}]\ket{{\rm RPA}}~,
 \end{equation}
where the RPA ground-state is determined as usual by the
requirement that $Q_{\nu}\ket{{\rm RPA}}=0$, we obtain the
following characteristic equation for RPA frequencies:
 \begin{eqnarray}\label{omega_eq}
 \Omega^4_{\nu}-\Omega^2_{\nu}\left(4\ep^2+\es^2+\frac{4g}{\ep}\right)+
 4\ep^2\left(\es^2+\frac{g}{\ep}\right)=0~.
 \end{eqnarray}
 The  solution is given by
 \[
  \Omega _\nu ^2=\frac \eta s+ \frac{8gs^2}{1-
  \frac{\displaystyle 4g}
  { \displaystyle \ep \left(\Omega _\nu ^2-4\ep ^2\right)} }~.
 \]
 Furthermore, using the relations in Eq.~(\ref{system}) it is easy to verify
 that the RPA frequencies obey the following condition:
 \begin{equation}
 (\Omega_1^2-\es^2)(\es^2-\Omega_2^2)=\frac{32g^2s^2}{\ep}\ ,
 \end{equation}
 and we shall assume that $\Omega^2_1>\Omega^2_2$.

 Using the orthonormalization condition of RPA states
 $\bra{\rm RPA}Q_{\nu}Q^+_{\nu'}\ket{\rm
 RPA}=\delta_{\nu\nu'}$~, one can find  the RPA amplitudes
 $x_{\nu},~y_{\nu},~u_{\nu},~v_{\nu}$ and express $\beta$ and
 ${\mathcal B}$ operators through  $Q_{\nu}$. However, as is seen
 from Eqs.~(\ref{H3}) and~(\ref{H4}) it is more convenient to use
 the inverse transformation $Q_{\nu}\ \to \ \beta,\ {\mathcal B}$
 in the following form:
 \begin{eqnarray}\label{amplit}
 \bp&=&\sqrt{\frac{2\ep}{\Omega_1}}\phi_1(Q_1^++Q_1)+\sqrt{\frac{2\ep}{\Omega_2}}
 \phi_2(Q_2^++Q_2), \nonumber\\
 \bm&=&\sqrt{\frac{\Omega_1}{2\ep}}\phi_1(Q_1^+-Q_1)+\sqrt{\frac{\Omega_2}{2\ep}}
 \phi_2(Q_2^+-Q_2),\nonumber\\
 (\beta^++\beta)&=&-\sqrt{\frac{\es}{\Omega_1}}\phi_2(Q_1^++Q_1)+
                    \sqrt{\frac{\es}{\Omega_2}}\phi_1(Q_2^++Q_2),\nonumber\\
 (\beta^+-\beta)&=&-\sqrt{\frac{\Omega_1}{\es}}\phi_2(Q_1^+-Q_1)+
                    \sqrt{\frac{\Omega_2}{\es}}\phi_1(Q_2^+-Q_2),
 \end{eqnarray}
where
 $\phi_1=\sqrt{\displaystyle{\frac{\Omega_1^2-\es^2}{\Omega_1^2-\Omega_2^2}}}$
 and
 $\phi_2=-\sqrt{\displaystyle{\frac{\es^2-\Omega_2^2}
{\Omega_1^2-\Omega_2^2}}}$~.

Now we are able to write down the $H_3$ and $H_4$ parts of the
Hamiltonian in terms of RPA bosons $Q_{\nu}$ and $Q^+_{\nu}$. For
the $H_3$ part one obtains
 \begin{equation}
   H_3 = H_3^{\pi\sigma}+H_3^{\sigma\sigma}~,
 \end{equation}
 with
 \begin{eqnarray}
   H_3^{\pi\sigma}=\frac{2}{\sqrt{K\ep}}
   \left\{\frac{\phi_1 R_1}{\sqrt{\Omega_1}}(Q_1^++Q_1)+
         \frac{\phi_2
 R_2}{\sqrt{\Omega_2}}(Q_2^++Q_2)\right\}(m-1)~,
 \end{eqnarray}
 \begin{eqnarray}
  H_3^{\sigma\sigma}&=&\sum_{\nu=1,2}A_{\nu 1}(Q_{\nu}^++Q_{\nu})^3+
  A_{\nu2}(Q_{\nu}^++Q_{\nu})^2(Q_{\nu'}^++Q_{\nu'})
  \nonumber\\&&+
   A_{\nu 3}(Q_{\nu}^+-Q_{\nu})^2(Q_{\nu}^++Q_{\nu})+
  A_{\nu 4}(Q_{\nu}^+-Q_{\nu})^2(Q_{\nu'}^++Q_{\nu'})
  \nonumber\\&&+
  A_{\nu5}(Q_{\nu}^+-Q_{\nu})(Q_{\nu}^++Q_{\nu})(Q_{\nu'}^++Q_{\nu})
  +({\rm h.c.})~,
 \end{eqnarray}
 where coefficients $A_{\nu i}$ ($i=1,\ldots, 5$) are
 \begin{eqnarray}
 A_{\nu 1}&=&\sqrt{\frac{\ep}{\Omega_{\nu}}}
 \frac{\phi_{\nu}}{2\Omega_{\nu}}\left(\frac{\phi^2_{\nu}
 R_{\nu}}{\sqrt{K}}+P_{\nu}\right)~,
 \nonumber\\
 A_{\nu 2}&=&\sqrt{\frac{\ep}{\Omega_{\nu'}}}
  \frac{\phi_{\nu'}}{2\Omega_{\nu}}\left(\frac{\phi^2_{\nu}
 R_{\nu'}}{\sqrt{K}}+
 \frac{2\phi^2_{\nu} R_{\nu}}{\sqrt{K}}+P_{\nu'}\right)~,
 \nonumber\\
 A_{\nu 3}&=&-\sqrt{\frac{\Omega_{\nu}}{\ep}}
 \frac{\phi^3_{\nu}}{8K}\left(\frac{R_{\nu}}{\sqrt{K}\ep}+\frac{P_{\nu}}{\omega}\right)~,
 \nonumber\\
 A_{\nu 4}&=&-\sqrt{\frac{\Omega_{\nu}}{\ep}}
 \frac{\phi^2_{\nu}\phi_{\nu'}\Omega_{\nu}}{8K\Omega_{\nu'}}\left(
 \frac{R_{\nu'}}{\sqrt{K}\ep}+\frac{P_{\nu'}}{\omega}\right)~,
 \nonumber\\
 A_{\nu 5}&=&-\sqrt{\frac{\Omega_{\nu'}}{\ep}}
 \frac{\phi^2_{\nu}\phi_{\phi'}}{4K}\left(\frac{R_{\nu}}{\sqrt{K}\ep}+
 \frac{P_{\nu}}{\omega}\right)~,\nonumber
 \end{eqnarray}
 and
 \[
 R_{\nu}=\frac{\Omega_{\nu}^2}{4}\sqrt{\frac{\ep}{\omega}}-\sqrt{K}\ep^2~,\quad
  P_{\nu}=\frac{\Omega_{\nu}^2}{4}-\ep^2~.
 \]
Note,  that $H_{3}^{\pi\sigma}$ is proportional to $(m-1)$ ($m$ --
is the pion number operator) whereas $H_{3}^{\sigma\sigma}$ only
includes the RPA boson operators $Q_{\nu}$,~$Q_{\nu}^+$.

Finally, the $H_4$ part of the Hamiltonian expansion reads:
 \begin{equation}
 H_4 = H_4^{\pi\pi}+H_4^{\pi\sigma}+H_4^{\sigma\sigma}
 \end{equation}
with
 \begin{eqnarray}
 H_4^{\pi\pi}&=&\frac1K\left(\ep d^2+\frac{g}{\ep\omega}\right)(m-1)^2~,
 \end{eqnarray}
 \begin{eqnarray}
 H_4^{\pi\sigma}&=&\frac{1}{2K}\sum_{\nu=1,2}\left\{\phi^2_{\nu}\left(\ep
 d^2+\frac{g}{\ep\omega}\right)\left(\frac{6\ep}{\Omega_{\nu}}+\frac{\Omega_{\nu}
 }{2\ep K}\right)+g\epw\right.
 \nonumber\\&&\times
 \left(\frac{1}{\Omega_{\nu}}+\frac{\Omega_{\nu}\phi^2_{\nu}}{4\ep\omega K}\right)
 +\left.\frac{32g^2s^2}{\omega\Omega_{\nu}(\Omega_{\nu}^2-\Omega_{\nu'}^2)}\right\}
 (2Q^+_{\nu}Q_{\nu}+1)(m-1)\nonumber\\&&
 +{\rm (off-diag.)}~,
 \end{eqnarray}
 \begin{eqnarray}
 H_4^{\sigma\sigma}&=&\sum_{\nu}
 \left\{\frac{\phi^2_{\nu}}{4K}\left(\frac{2\ep}{\Omega_{\nu}}+\frac{\Omega_\nu}{
 2\ep K}\right){\mathcal F}^+_{\nu}+
 \frac{\phi^2_{\nu}}{8K}\left(\frac{2\ep}{\Omega_{\nu}}-\frac{\Omega_\nu}{2\ep
 K}\right){\mathcal F}^-_{\nu}+ {\mathcal
 R}_{\nu}\right\}
 \nonumber\\&&\times(Q^+_{\nu}Q^+_{\nu}Q_{\nu}Q_{\nu}-2Q^+_{\nu}Q_{\nu})
 \nonumber\\&&+
 \frac12\left\{\frac{\phi^2_2}{4K}\left(\frac{2\ep}{\Omega_2}+\frac{\Omega_2}{2\ep
 K}\right){\mathcal F}^+_1+
        \frac{\phi^2_1}{4K}\left(\frac{2\ep}{\Omega_1}+\frac{\Omega_1}{2\ep
 K}\right){\mathcal F}^+_2+{\mathcal
        P}\right\}\times\nonumber\\&&\times
     (2Q^+_1Q^+_2Q_1Q_2-Q^+_1Q_1-Q^+_2Q_2)+({\rm off-diag.})~,
  \end{eqnarray}
 where
 \begin{eqnarray}
  {\mathcal F}_{\nu}^{\pm}&=&
   \phi^2_{\nu}\left(\ep d^2+\frac{g}{\ep\omega}\right)
   \left(\frac{10\ep}{\Omega_{\nu}}\pm\frac{\Omega_{\nu}}{2K\ep}\right)+
   \frac{64g^2s^2}{\omega\Omega_{\nu}(\Omega^2_{\nu}-\Omega^2_{\nu'})}\nonumber\\&&+
  2g\epw\left(\frac{1}{\Omega_{\nu}}\pm\frac{\phi^2_{\nu}\Omega_{\nu}}{4K\ep\omega}\right)~,
   \nonumber\\
   {\mathcal R}_{\nu}&=&\frac{g}{2}\left(\frac{3}{\Omega_{\nu}^2}+
   \frac{\phi_{\nu}^2}{2K\ep\omega}+
   \frac{3\phi^4_{\nu}\Omega^2_{\nu}}{16K^2\ep^2\omega^2}\right)~,
   \nonumber\\
   {\mathcal P}&=&\frac{\phi^2_1\phi^2_2}{K}\left\{
   \left(\ep
  d^2+\frac{g}{\ep\omega}\right)\left(\frac{20\ep^2}{\Omega_1\Omega_2}+
   \frac{\Omega_1\Omega_2}{4K^2\ep^2}\right)-
   \frac{2\ep^2(\Omega^2_1+\Omega^2_2-2\es^2)}{\Omega_1\Omega_2\omega}\right.\nonumber\\&&+
   \left. g\frac{\Omega_1\Omega_2}{4K\ep^2\omega^2}+
   g\epw\frac{\Omega_1\Omega_2}{4K^2\ep^2\omega}\right\}\nonumber\\&&+
   2g\left(\frac{1}{\Omega_1}+\frac{\phi^2_1\Omega_1}{4K\ep\omega}\right)
 \left(\frac{1}{\Omega_2}+\frac{\phi^2_2\Omega_2}{4K\ep\omega}\right)~.\nonumber
  \end{eqnarray}

 %%%%%%%%%%%%%%%%%%%%%%%%%%%%%%%%%%%%%%%%%
\section{}
%\renewcommand{\theequation}{C.\arabic{equation}}
%\setcounter{equation}{0}
%%%%%%%%%%%%%%%%%%%%%%%%%%%%%%%%%%%%%%%%%
In the present Appendix the expression of the operator $iS_d$ as
well as the terms of the transformed expansion of the Hamiltonian
(\ref{hdiag}) are given.

In terms of RPA operators, the requirement (\ref{cond}) to
eliminate the off-diagonal terms of order $1/\sqrt{N}$ reads
  \begin{equation}\label{cond2}
  [H_2,iS]_d=\sum_{\nu}\Omega_{\nu}[Q_{\nu}^+Q_{\nu},iS]_d=-(H_3)_d~.
  \end{equation}
 The solution of Eq.~(\ref{cond2}) can be found in the following form:
  \begin{equation}\label{gener}
  iS_d=iS_d^{\pi\sigma}+iS_d^{\sigma\sigma}
  \end{equation}
  with
  \begin{eqnarray}\label{gen_coef}
  iS_d^{\pi\sigma}&=&\sum_{\nu}\lambda_{\nu}(Q_{\nu}^+-Q_{\nu})_d(m-1)_d~,\nonumber\\
  iS_d^{\sigma\sigma}&=&
  \sum_{\nu}\alpha_{\nu}(Q_{\nu}^+-Q_{\nu})_d^3+
            \beta_{\nu}(Q_{\nu}^+-Q_{\nu})_d^2(Q_{\nu'}^+-Q_{\nu'})_d\nonumber\\&&+
            \gamma_{\nu}(Q_{\nu}^+-Q_{\nu})_d(Q_{\nu}^++Q_{\nu})_d^2+
            \delta_{\nu}(Q_{\nu'}^+-Q_{\nu'})_d(Q_{\nu}^++Q_{\nu})_d^2\nonumber\\&&+
            \rho_{\nu}(Q_{\nu}^+-Q_{\nu})_d(Q_{\nu}^++Q_{\nu})_d(Q_{\nu'}^+-Q_{\nu'})_d-
            {\rm h.c.}\nonumber
  \end{eqnarray}
 Inserting these expressions into Eq.~(\ref{cond2}) we immediately
 find
 \begin{equation}\label{coef1}
 \lambda_{\nu}=-\frac{2}{\sqrt{K\ep}}\frac{\phi_{\nu}R_{\nu}}{\Omega^{3/2}_{\nu}}~,\quad
 \alpha_{\nu}=\frac{2A_{\nu 1}-A_{\nu 3}}{3\Omega_{\nu}}~,\quad
 \gamma_{\nu}=-\frac{A_{\nu 1}}{\Omega_{\nu}}~.
 \end{equation}
 The other three coefficients in Eq.~(\ref{gen_coef}) are the
 solutions of a linear system of equations:
 \begin{eqnarray}
 \left(\begin{array}{ccc}
                        0         &   \Omega_{\nu'}& \Omega_{\nu}  \\
                   2\Omega_{\nu}  &  2\Omega_{\nu} & \Omega_{\nu'} \\
                    \Omega_{\nu'} &     0          & \Omega_{\nu}  \
 \end{array}\right)
  \left(\begin{array}{c}
                        \beta_{\nu}\\
                        \delta_{\nu}\\
                        \rho_{\nu}\
                                        \end{array}\right)=
 -\left(\begin{array}{c}
                        A_{\nu 2}\\
                        A_{\nu 5}\\
                        A_{\nu 4} \
                                        \end{array}\right)~,\nonumber
 \end{eqnarray}
which gives
 \begin{eqnarray}\label{coef2}
 \beta_{\nu}&=&
 \frac{2A_{\nu2}\Omega_{\nu}^2-A_{\nu 4}(2\Omega_{\nu}^2-\Omega_{\nu'}^2)-A_{\nu
 5}\Omega_1\Omega_2}
 {\Omega_{\nu'}(4\Omega_{\nu}^2-\Omega_{\nu'}^2)}~,\nonumber\\
  \delta_{\nu}&=&
 \frac{2A_{\nu 4}\Omega_{\nu}^2-A_{\nu
 2}(2\Omega_{\nu}^2-\Omega_{\nu'}^2)-A_{\nu 5}\Omega_1\Omega_2}
 {\Omega_{\nu'}(4\Omega_{\nu}^2-\Omega_{\nu'}^2)}~,\nonumber\\
 \rho_{\nu}&=&
 \frac{A_{\nu 5}\Omega_{\nu'}-2\Omega_{\nu}(A_{\nu 2}+A_{\nu 4})}
 {4\Omega_{\nu}^2-\Omega_{\nu'}^2}~.
 \end{eqnarray}

With the condition (\ref{cond2}) one can eliminate the double
commutators from Eq.~(\ref{hdiag}) to obtain the following
expression for the Hamiltonian $H$:
 \begin{eqnarray}\label{hexpres}
 H&=&NH_0+(H_2)_d+\frac{1}{2N}\Bigl\{[(H_3^{\pi\sigma}+H_3^{\sigma\sigma}),(iS^{\pi
 \sigma}+iS^{\sigma\sigma})]\nonumber\\&&+
 2(H_4^{\pi\pi}+H_4^{\pi\sigma}+H_4^{\sigma\sigma})\Bigl\}_d+\cdots~.
 \end{eqnarray}
It contains all corrections to RPA which are of the order $1/N$.
These corrections can be calculated by evaluating the expectation
value of $H$. Of course, one could perform an additional unitary
transformation to remove the remaining off-diagonal terms which
are still present to order $1/N$. This is, however, only required
if one is interested in energies to order $1/N^2$. Since those are
not needed for the present purpose all off-diagonal terms can be
disregarded.

Explicitly, the commutators appearing in Eq.~(\ref{hexpres}) are
given by
  \begin{eqnarray*}
  [ H^{\pi\sigma}_{3},iS^{\pi\sigma} ]_d  = -\frac{8}{K\ep}\left(
  \frac{\phi_1^2R_1^2}{\Omega_1^2}+
  \frac{\phi_2^2R_2^2}{\Omega_2^2}\right)(m-1)^2_d~,
  \end{eqnarray*}
   \begin{eqnarray}
  \left[H_{3}^{\pi\sigma},iS^{\sigma\sigma}\right]_d&+&[H_{3}^{\sigma\sigma},iS^{\pi
  \sigma}]_d=\nonumber\\&&
  -\frac{16}{\sqrt{K\ep}}\sum_{\nu}
  \left(\frac{R_{\nu}\phi_{\nu}(3A_{\nu1}-A_{\nu3})}{\Omega_\nu^{3/2}}+
  \frac{R_{\nu'}\phi_{\nu'}(A_{\nu2}-A_{\nu4})}{\Omega_{\nu'}^{3/2}}\right)
        \nonumber\\&&\times
   (2Q_\nu^+Q_\nu+1)_d(m-1)_d+{\rm (off-diag.)}~,
   \nonumber
   \end{eqnarray}
   \begin{eqnarray}
   \left[H^{\sigma\sigma}_{3},iS^{\sigma\sigma}\right]_d&=&16\sum_{\nu}\Bigl\{
   9A_{\nu1}(\gamma_{\nu}-\alpha_{\nu})+A_{\nu2}(3\delta_{\nu}-\beta_{\nu})+
   3A_{\nu3}(3\alpha_{\nu}+\gamma_{\nu})
   \nonumber\\&&
   +A_{\nu4}(3\beta_{\nu}-\delta_{\nu})+A_{\nu5}\rho_{\nu}\Bigl\}
   (Q^+_{\nu}Q^+_{\nu}Q_{\nu}Q_{\nu}-2Q^+_{\nu}Q_{\nu})_d\nonumber\\&&+
   16\Bigl\{(\beta_2-\delta_2)(A_{13}-3A_{11})+(\gamma_2-3\alpha_2)(A_{12}-A_{14})
   \nonumber\\&&+
    2\rho_1(A_{12}+A_{14})+2A_{15}(\beta_1+\delta_1)+
    (\beta_1-\delta_1)(A_{23}-3A_{21})
   \nonumber\\&&
   +(\gamma_1-3\alpha_1)(A_{22}-A_{24})+
   2\rho_2(A_{22}+A_{24})+2A_{25}(\beta_2+\delta_2)\Bigl\}\nonumber\\&&\times
             (2Q^+_1Q^+_2Q_2Q_1-Q^+_1Q_1-Q^+_2Q_2)_d+({\rm
             off-diag.})~.\nonumber
  \end{eqnarray}
 With these commutators we can write the next-to-leading order
 corrections  in the following form:
 \begin{eqnarray}\label{correct}
   \frac{1}{2}[H^{\pi\sigma}_{3},iS^{\pi\sigma}]_d+H^{\pi\pi}_{4d}&=&
   \frac{1}{2}
   V_{\pi\pi\to\pi\pi}(m-1)^2_d\nonumber\\&=&\frac{1}{2}
   V_{\pi\pi\to\pi\pi}(\sum_{i,j}\alpha^+_i\alpha^+_j\alpha_j\alpha_i-
   \sum_i\alpha_i^+\alpha_j+1)~,
  \nonumber\\
  \frac{1}{2}\left([H_{3}^{\pi\sigma},iS^{\sigma\sigma}]_d+
 [H_{3}^{\sigma\sigma},iS^{\pi\sigma}]_d\right)&+&H^{\pi\sigma}_{4d}\nonumber\\&=&
 \frac{1}{2}\sum_{\nu}V_{\pi\sigma_{\nu}\to\pi\sigma_{\nu}}(2Q_\nu^+Q_\nu+1)_d(m-1)_d\nonumber\\&&+
  {(\rm off-diag.)}~,
   \nonumber\\
  \frac{1}{2}[H^{\sigma\sigma}_{3},iS^{\sigma\sigma}]_d+H^{\sigma\sigma}_{4d}&=&
  \frac{1}{2}\sum_{\nu}V_{\sigma_\nu\sigma_\nu\to\sigma_\nu\sigma_{\nu}}
 (Q^+_{\nu}Q^+_{\nu}Q_{\nu}Q_{\nu}-2Q^+_{\nu}Q_{\nu})_d\nonumber\\&&+
             \frac{1}{2} V_{\sigma_1\sigma_2\to\sigma_1\sigma_2}
             (2Q^+_1Q^+_2Q_2Q_1-\sum_{\nu}Q^+_{\nu}Q_{\nu})_d\nonumber\\&&+
             {(\rm off-diag.)}~,\nonumber
  \end{eqnarray}
 where
  \begin{eqnarray*}%\label{matr_el}
  V_{\pi\pi\to\pi\pi}&=&\frac{8g\ep^2}{\Omega_1^2\Omega_2^2}~;\\
  V_{\pi\sigma_{\nu}\to\pi\sigma_{\nu}}&=&
  \frac{8g\ep^2}{\Omega_1^2\Omega_2^2}
 \left\{\frac{\ep}{\Omega_\nu}+
 \frac{(\Omega_\nu^2-\ep^2)(4\ep^2+\Omega_{\nu'}^2)}
 {\Omega_\nu\ep(\Omega_\nu^2-\Omega_{\nu'}^2)}\right\}~;\\
 \end{eqnarray*}
\begin{eqnarray}\label{matr_el}
 V_{\sigma_\nu\sigma_\nu\to\sigma_\nu\sigma_{\nu}}&=
 & \frac{6g\ep^4}{\Omega_{\nu}^4\Omega_{\nu'}^2}+
 \frac{6g(\Omega_{\nu}^2-\ep^2)(8\ep^4+2\ep^2\Omega_{\nu'}^2-\Omega_1^2\Omega_2^2
 )}{
 \Omega_{\nu}^4\Omega_{\nu'}^2(\Omega_{\nu}^2-\Omega_{\nu'})}\nonumber\\&&+
 \frac{2g(4\ep^2-\Omega^2_{\nu'})}
 {3\Omega_{\nu}^4\Omega_{\nu'}^2(\Omega_1^2-\Omega_2^2)^2}\nonumber\\&&\times
 \left\{\Omega_{\nu}^4(17\ep^2-10\Omega_{\nu'}^2)+(\ep^4-20\ep^2\Omega_{\nu}^2)
 (20\ep^2+7\Omega^2_{\nu'})\right\}\nonumber\\&&+
 \frac{2g(\Omega_{\nu}^2-4\ep^2)^2(4\ep^2-\Omega^2_{\nu'})(\Omega_{\nu'}-\ep^2)}
 {3\Omega_1^2\Omega_2^2(4\Omega_{\nu}^2-\Omega_{\nu'}^2)(\Omega_1^2-\Omega_2^2)^2}~;
 \nonumber\\
 V_{\sigma_1\sigma_2\to\sigma_2\sigma_1}&=&\sum_{\nu}
 \frac{4g(4\ep^2-\Omega_{\nu}^2)(\Omega_{\nu'}^2-\ep^2)}{
       3\Omega_1^3\Omega_2^3(\Omega_1^2-\Omega_2^2)^2}\nonumber\\&&\times
 \left\{2(4\ep^2\Omega_{\nu'}^2-5\Omega_1^2\Omega_2^2-10\ep^2\Omega^2_{\nu}+10\ep^4)\nonumber\right.\\&&+\left.
 \frac{(4\ep^2-\Omega_{\nu}^2)(4\ep^2-\Omega_{\nu'}^2)\Omega_{\nu'}^2}{4\Omega_{\nu}^2-\Omega_{\nu'}^2}
 \right\}~.
  \end{eqnarray}
Thus there are three sources of the residual interaction:
pion-pion, pion-sigma and sigma-sigma couplings. Furthermore, we
can evaluate the higher-order corrections to the pion mass as
  \begin{eqnarray}\label{pion_pr}
  \ep'=-\frac{1}{2}\Bigl\{V_{\pi\pi\to\pi\pi}-\sum_{\nu=1,\,2}V_{\pi\sigma_\nu\to\pi\sigma_\nu}\Bigl\}~;
  \end{eqnarray}
  and to the sigma-boson mass as
  \begin{eqnarray}\label{sigma_pr}
  \Omega'_{\nu}=-\frac{1}{2}\left\{2V_{\pi\sigma_\nu\to\pi\sigma_\nu}+
2V_{\sigma_{\nu}\sigma_{\nu}\to\sigma_{\nu}\sigma_{\nu}}+V_{\sigma_1\sigma_2\to
 \sigma_2\sigma_1}\right\}~.
  \end{eqnarray}

 \end{document}